\documentclass[11pt,twoside]{article}

\usepackage{asp2014}
\aspSuppressVolSlug
\resetcounters

\begin{document}
\title{Implementing SIAv2 Over Rubin Observatory's Data Butler}
\author{Tim~Jenness,$^1$ Stelios~Voutsinas,$^1$ Gregory~P.~Dubois-Felsmann,$^2$ and Andrei~Salnikov$^3$}
\affil{$^1$Vera C.\ Rubin Observatory Project Office, 950 N.\ Cherry Ave., Tucson, AZ  85719, USA}
\affil{$^2$Caltech/IPAC, California Institute of Technology, MS 100-22, Pasadena, CA 91125-2200, USA}
\affil{$^3$SLAC National Accelerator Laboratory,  2575 Sand Hill Rd., Menlo Park, CA 94025, USA}
\paperauthor{Tim~Jenness}{}{0000-0001-5982-167X}{Vera C.\ Rubin Observatory Project Office}{}{Tucson}{AZ}{85719}{USA}
\paperauthor{Stelios~Voutsinas}{}{0009-0003-4290-2942}{Vera C.\ Rubin Observatory Project Office}{}{Tucson}{AZ}{85719}{USA}
\paperauthor{Gregory~P.~Dubois-Felsmann}{}{0000-0003-1598-6979}{Caltech/IPAC}{}{Pasadena}{CA}{91125-2200}{USA}
\paperauthor{Andrei~Salnikov}{}{0000-0002-3623-0161}{SLAC National Accelerator Laboratory}{}{Menlo Park}{CA}{94025}{USA}
% Yes they said to have these index commands commented out.
%\aindex{Jenness,~T.}
%\aindex{Voutsinas,~S.}
%\aindex{Dubois-Felsmann,~G.~P.}
%\aindex{Salnikov,~A.}

% This can write metadata into the PDF.
% Update keywords and author information as necessary.
\hypersetup{
    pdftitle={Implementing SIAv2 Over Rubin Observatory's Data Butler},
    pdfauthor={jennesst},
    pdfkeywords={}
}

\begin{abstract}
The IVOA Simple Image Access version 2 protocol defines an easy way to provide community access to a collection of data.
At the Vera C.\ Rubin Observatory we currently enable ObsTAP access to our data holdings via an ObsCore export or view of our Data Butler repositories.
This approach comes with some deployment constraints, such as requiring pgsphere and compatibility with our CADC TAP implementation, so recently we decided to see whether we could instead provide an SIAv2 service that talks directly to our Data Butler.
Here we describe our motivation, implementation strategies, and current deployment status, as well as discussing some metadata mismatches between the Butler data models and SIAv2.
\end{abstract}

% These lines show examples of subject index entries. At this stage these have to commented
% out, and need to be on separate lines. Eventually, they will be automatically uncommented
% and used to generate entries in the Subject Index at the end of the Proceedings volume.
%\ssindex{Virtual Observatory (VO)!standards!Simple Image Access}
%\ssindex{observatories!ground-based!Rubin}

% These lines show examples of ASCL index entries. At this stage these have to commented
% out, and need to be on separate lines. Eventually, they will be automatically uncommented
% and used to generate entries in the ASCL Index at the end of the Proceedings volume.
% The ascl.py command will scan your paper on possible code names.
% Don't leave these in! - replace them with ones relevant to your paper.
%ooindex{FOOBAR, ascl:1101.010}

\section{Motivation}

The Rubin Data Butler \citep{2022SPIE12189E..11J} consists of a metadata registry and a file datastore.
The registry contains sufficient information to construct ObsCore records.
Previously it was possible to provide an ObsCore table using two methods: export the records as CSV or Parquet files and load them into a static database, or provide live syncing to an ObsCore table using hooks in the registry backend \citep{DMTN-236}.

Both these approaches have been implemented by us and are the only way we had to provide an ObsTAP service \citep{2017ivoa.spec.0509L}.
However, there are downsides.
Whilst static export works fine for formal data releases where it can be integrated into our high performance Qserv database \citep{C15_adassxxxii}, it is not suitable for evolving datasets such as those from the nightly prompt products.
``Live'' ObsCore works but requires that the deployment has pgsphere available.
Also, if the configuration changes (something that can happen frequently during early operations) the entire table needs to be rebuilt.

These two issues suggested that there would be a benefit from providing a simpler, yet standardized, query layer directly atop the Butler.
The IVOA's Simple Image Access version 2 \citep[SIAv2;][]{2015ivoa.spec.1223D} was the obvious choice for this.
Interfacing directly to the Butler provides much more flexibility since configuration changes can be picked up with a simple restart of the service and it can work immediately with any Butler repository.

\section{Implementation}

\articlefigure{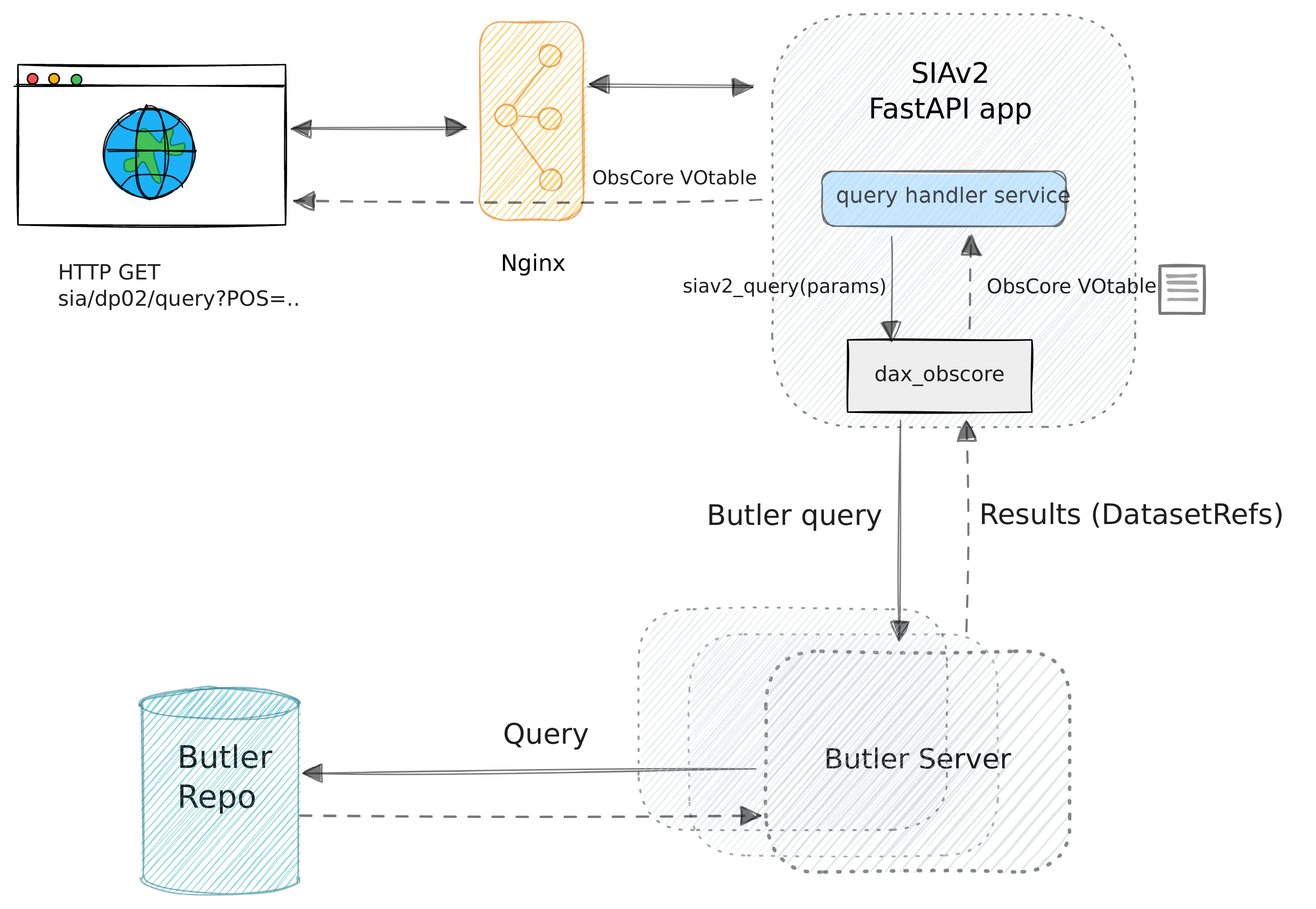}{fig:arch}{Architecture diagram showing how the SIAv2 service interfaces with the Butler.}

We used a layered approach to implementation as shown in Fig.\ \ref{fig:arch}.
The service itself \citep{SQR-095} is written in Python and FastAPI using Rubin's standard internal development platform Phalanx.\footnote{\url{https://phalanx.lsst.io}}
This provides us with our standard authentication layer and deployment capabilities.

The service takes the raw SIAv2 parameters and passes them, along with the SIAv2 configuration specific for the attached Butler repository, to the \texttt{dax\_obscore}\footnote{\url{https://github.com/lsst-dm/dax_obscore}} library which parses the parameters, converts them to Butler queries, runs the queries, and returns the standardized results as an Astropy VOTable that can be packaged up by the service and returned.
The resulting table schema is defined by a Felis data model for consistency \citep{C702_adassxxxiv}.
The service works transparently with both the original ``direct'' Butler and the new client/server remote Butler \citep{2024SPIE13101E..3GJ}.

Separating the service layer from the SIAv2 query handling allows us to test the SIA-to-Butler queries much more easily.
Additionally, the \texttt{dax\_obscore} package, which can be installed from PyPI, provides a command-line interface to make it easier for people to learn the SIAv2 parameter system and experiment with queries locally.

\section{Current Status}

A service has been deployed on the Rubin Science Platform and is ready to be used with commissioning data.
The \texttt{dax\_obscore} package supports the following SIAv2 query parameters at this time:
\texttt{MAXREC},
\texttt{INSTRUMENT},
\texttt{POS},
\texttt{TIME},
\texttt{BAND},
\texttt{EXPTIME},
\texttt{CALIB}.

Butler itself has native support for region and time queries.
Support for
\texttt{ID},
\texttt{TARGET},
\texttt{FACILITY},
and \texttt{COLLECTION} are coming soon.
For the \texttt{FACILITY} keyword we intend to match the AAS Facility names of ``Rubin:Simonyi'' for the Simonyi Survey Telescope and ``Rubin:1.2m'' for the Rubin Auxiliary Telescope.
The Butler does not have a concept of facility but the correct value will be derived from the instrument name.
Butler does have a concept of target, but in general, the value of the field is not of interest, since it will usually refer to a field name generated by the survey scheduler.

\section{Model Mismatches}

The Butler data model as used by Rubin Observatory results in some implementation difficulties.

\subsection{Instrument for coadds}

Currently, as defined by the Rubin Science Pipelines team, the Butler registry does not have an instrument associated with co-adds.
The instrument is recorded in the datasets and in the collection names but when the pipelines were initially defined it was deemed to be unimportant to include the instrument information in the dataset type definition itself.
This causes difficulties in SIAv2 in a Butler repository where LATISS and LSSTCam data are both available since there is no way to know from the query system which is which.
Currently it is deemed to be too disruptive to modify the pipeline connections to propagate the instrument into the co-adds Butler dataset definition.
Once we implement full provenance tracking it should be possible to determine the instrument using the provenance to determine the original datasets that went into the co-add.

\subsection{Exposure time for co-adds}

Individual observations include exposure times in the Butler registry metadata, but this information is not available to co-adds.
This is because the median exposure time for a co-add is a derived quantity that is not known when the Butler coordinate space is defined.
Storing derived metadata, which is similar to calculating the seeing and then allowing a query based on seeing, is on the future development roadmap.

\subsection{Observing dates for co-adds}

Butler knows the observing dates of individual observations, but for a co-add this information is lost.
In the future when a full Butler provenance system is implemented it may be possible to derive the date range for co-adds, but it is not possible at this time.

\subsection{Dataset Types}

Butler makes extensive use of what we call ``dataset types'' which define each product type in a pipeline.
Examples can be \texttt{visit\_image}, \texttt{difference\_image}, and \texttt{deep\_coadd}.
Currently there is no standardized way in SIAv2 for a query to specify these.
We are looking at adding an extension, potentially using a \texttt{DPSUBTYPE} query parameter that will map directly to the Butler dataset type, possibly with an \texttt{lsst} prefix.

\section{Conclusions}

Implementing SIAv2 over the Data Butler was a relatively simple process.
Separating the development into a standalone SIAv2 parameters handler in \texttt{dax\_obscore} with a service that forwards parameters to this layer allowed for parallel development.
An additional advantage is that there is now a command-line tool that can be used to perform SIAv2-style queries directly on any Butler repository.
Some data model mismatches are causing difficulties, but we hope to fix most of those issues with the additional support for querying on derived metadata and provenance.

\acknowledgments This material or work is supported in part by the National Science Foundation through Cooperative Agreement AST-1258333 and Cooperative Support Agreement AST1836783 managed by the Association of Universities for Research in Astronomy (AURA), and the Department of Energy under Contract No.\ DE-AC02-76SF00515 with the SLAC National Accelerator Laboratory managed by Stanford University.

\bibliography{P920}

\end{document}